\documentclass[%
 reprint,
superscriptaddress,
 amsmath,amssymb,
 aps,
 pre,
]{revtex4-2}

\usepackage{graphicx}
\usepackage{dcolumn}
\usepackage{bm}
\usepackage{braket} 
\usepackage{siunitx}
\DeclareSIUnit\angstrom{\text {Å}}
\DeclareSIUnit\bohr{\text {\ensuremath {a}}_{0}}
\usepackage{booktabs}
\usepackage{xcolor}
\begin{document}
\preprint{APS/123-QED}

\title{Assessing the Projector Augmented-Wave Method for Stopping Power Calculations}

\author{Bryn Lloyd}
 \email{bryn.lloyd@physics.ox.ac.uk}
\affiliation{Department of Physics, University of Oxford, OX1 3PU, United Kingdom}

\author{Dirk O. Gericke}
\affiliation{Centre for Fusion, Space and Astrophysics, Department of Physics, University of Warwick, Coventry CV4 7AL, United Kingdom}

\author{Gilles Rodway-Gant}
\affiliation{Alpha Ring International Limited, 5 Harris Ct Building B, Monterey, California 93940, USA}
\affiliation{Cavendish Laboratory, Department of Physics, University of Cambridge, JJ Thomson Ave., Cambridge CB3 0US, United Kingdom}

\author{Gianluca Gregori}%
\affiliation{Department of Physics, University of Oxford, OX1 3PU, United Kingdom}

\date{\today}

\begin{abstract}

The stopping power of charged particles is investigated using time-dependent density functional theory (TDDFT). Such simulations are made possible by recent advances in computational resources and numerical implementations of this first-principles method. In practice, DFT simulations widely employ the projector augmented-wave (PAW) method to approximate all-electron behaviour, but the implications of the PAW approximation for non-adiabatic stopping simulations remain insufficiently explored. Here, the suitability of the PAW method for stopping power simulations is evaluated. A workflow for generating and selecting PAW datasets tailored to these simulations is developed, enabling systematic optimisation of augmentation radii and projector constructions. The approach is applied to proton stopping in FCC aluminium, demonstrating how dataset design influences stopping predictions, and enabling an investigation of crystal channelling effects on charged-particle transport.
\end{abstract}

\maketitle

\section{\label{sec:introduction}Introduction}

Stopping power quantifies the force on 
an energetic charged particle (projectile) travelling through matter, which is generally expressed as the energy loss per unit distance travelled. Accurate knowledge of stopping powers is important in a range of contexts, including ion implantation in semiconductor fabrication \cite{chen_improving_2024}, radiation damage in irradiated materials \cite{sand_heavy_2019}, and medical ion-beam therapy \cite{matias_efficient_2024}. Energy production via nuclear fusion is an other important application. Here, the stopping of fusion-born $\alpha$-particles governs the distribution of the internally produced energy and the propagation of fusion burn \cite{hu_review_2024}.

At sufficiently low projectile energies, nuclear stopping, the transfer of energy to host nuclei through collisions with the projectile, dominates energy loss. This process can be modelled using approaches such as molecular dynamics or, where the dynamics can be approximated as a sequence of independent collisions, binary collision models \cite{robinsonComputerSimulationAtomicdisplacement1974}. For projectiles with velocities faster than typical ion thermal velocities, stopping power due to the target electrons becomes the dominant contribution. As such, the modelling of electronic stopping is the primary focus of this work. Excitations of electrons are more challenging to model than classical collisions, and in nearly a century of research, various models have been proposed. One of the earliest models is the Bethe-Bloch model, in which electronic stopping arises from interactions between the projectile and atomic electrons, which are assumed to be bound to infinitely massive nuclei \cite{bethe_zur_1930, blochZurBremsungRasch1933}. In the decades since its development, various correction terms have been added \cite{fanoPenetrationProtonsAlpha1963}. For metals and plasmas, where contributions from free electrons are significant, Lindhard developed a fundamentally different approach based on the linear response of an electron gas \cite{Lindhard1954}. Subsequent models have extended this framework to describe electronic stopping in dense plasmas \cite{deutsch_inertial_1986, peter_energy_1991,  pitarke_z13_1993, gericke_stopping_1996}.

A feature of many stopping models is the restriction to projectiles with a fixed charge-state. The Bethe-Bloch model, generally considered accurate at high velocities, is a typical example. In practice, stopping powers for many applications are obtained using semi-empirical tools such as SRIM (Stopping and Range of Ions in Matter) \cite{ziegler_srim_2010, ziegler_srim_2015}. SRIM combines the Bethe-Bloch formalism with empirical corrections derived from experimental data and fitted stopping powers at low energies where the high-velocity assumptions of the Bethe-Bloch model break down. A further consequence of this approach is that materials are assumed to be homogeneous. While these approaches provide reliable stopping powers for many materials and energies, their semi-empirical nature limits their predictive capability for systems where experimental data is scarce or where atomic-scale effects, such as crystal orientation are important.

While the framework of density functional theory (DFT) and time-dependent density functional theory (TDDFT) has long been established, their application to charged-particle stopping in solids has only recently become computationally feasible. Advances in high-performance computing and the efficient implementation of real-time propagation schemes, together with the widespread adoption of pseudopotential-based approaches have enabled routine first-principles simulations of electronic stopping using TDDFT. TDDFT naturally captures charge transfer and dynamic screening around the projectile through the time-dependent evolution of the electronic density. In practice, rapidly varying wavefunctions of tightly bound core electrons render all-electron calculations computationally demanding, often necessitating the use of a pseudisation scheme. Among these, the projector augmented-wave (PAW) method, originally formulated by Bl\"ochl \cite{blochlProjectorAugmentedwaveMethod1994}, has become widely adopted in DFT codes for its ability to systematically approach all-electron accuracy \cite{lejaeghere_reproducibility_2016}. The central advantage of the PAW method over conventional pseudopotential approaches is that it employs smoothed wavefunctions that retain a transformation to reconstruct the corresponding all-electron wavefunctions. Despite the widespread use of PAW in DFT, the implications of the PAW approximation for TDDFT-based simulations of electronic stopping have not yet been systematically investigated.

In this work, the suitability of the PAW method for calculating electronic stopping powers using TDDFT-based Ehrenfest dynamics simulations is evaluated. To this end, a systematic procedure for generating and assessing PAW datasets suitable for TDDFT stopping power calculations is developed. Finally, this framework is applied to both off-channelling and channelling trajectories in FCC aluminium, enabling comparison to SRIM and an investigation of the influence of crystal structure on charged-particle transport.

\section{\label{sec:theory}Theory}
\subsection{\label{subsec:ehrenfest_dynamics}Ehrenfest Dynamics}

Real-time TDDFT simulations are performed within the framework of Ehrenfest dynamics, in which electronic wavefunctions evolve in the time-dependent potential generated by the classical nuclear trajectories, while the nuclei experience forces derived from the instantaneous electronic density.

The projectile is treated as a nucleus moving through the simulation box, inducing a time-dependent electronic response. Energy may therefore be transferred from the projectile both to electronic excitations and, through the forces acting on the classical nuclei, to the motion of the host atoms. Electronic and nuclear contributions to the projectile energy loss are both contained within the coupled electron-nuclear dynamics. The Kohn-Sham wavefunctions, $\ket{\psi_n}$, are propagated according to the time-dependent Kohn–Sham equations

\begin{equation}
i \hbar \frac{\partial \ket{\psi_n}}{\partial t}
=
\hat{\mathcal{H}}[\mathbf{R}(t)] \ket{\psi_n},
\end{equation}
\noindent
where the Kohn-Sham Hamiltonian depends on the instantaneous nuclear positions, $\mathbf{R}(t)$,

\begin{equation}
\hat{\mathcal{H}}[\mathbf{R}(t)]
=
-\frac{\hbar^2}{2 m_e} \nabla^2 + V_{\rm ext}[\mathbf{R}(t)] + V_H[\rho] + V_{\rm XC}[\rho].
\end{equation}
\noindent
Here, the external potential $V_{ext}$ depends on the instantaneous nuclear positions. $V_H$ is the Hartree potential and $V_{\rm XC}$ is the exchange-correlation potential, both of which are functionals of the electronic density, $\rho$.

The nuclei evolve according to classical equations of motion,

\begin{equation}
M_a \ddot{\mathbf{R}}_a
=
-
\sum_n f_n
\left\langle \psi_n(t)
\middle|
\frac{\partial \hat{\mathcal{H}}[\mathbf{R}(t)]}{\partial \mathbf{R}_a}
\middle|
\psi_n(t)
\right\rangle,
\end{equation}
\noindent
where $M_a$ and $\ddot{\mathbf{R}}_a$ are the mass and acceleration of nucleus $a$, respectively. In the case where the electronic system is in its ground-state, these forces reduce to Hellman-Feynman forces. $f_n$ is the occupation number of Kohn-Sham state $n$, determined from the ground-state within the finite-temperature Mermin formalism \cite{merminThermalPropertiesInhomogeneous1965}.

\subsection{\label{subsec:projector_augmented_wave_formalism}Projector Augmented-Wave Formalism}
The projector augmented-wave method defines a linear transformation between pseudo and all-electron quantities. This transformation is given by

\begin{equation}
\label{eqn:paw_transformation}
\begin{aligned}
\ket{\psi_n}
&= \hat{\mathcal{T}} \ket{\tilde{\psi_n}} \\
&= \left(
    1 + \sum_a \sum_i
    \left( \ket{\phi_i^a} - \ket{\tilde{\phi}_i^a} \right)
    \bra{\tilde{p}_i^a}
\right) \ket{\tilde{\psi_n}}.
\end{aligned}
\end{equation}
\noindent
$\ket{\psi_n}$ and $\ket{\tilde{\psi}_n}$ are the all-electron and pseudo-wavefunctions, respectively. $\ket{\phi_i^a}$ and $\ket{\tilde{\phi_i^a}}$ are the all-electron and pseudo partial-waves centred on atom $a$. $\ket{\tilde{p}_i^a}$ are the corresponding projector functions. $n$ indicates a Kohn-Sham orbital, and $i$ is a multi-index used to label partial-wave and projector channels, including principal and orbital angular momentum quantum numbers. Whilst this transformation is exact for infinitely-many projectors, in practice a small number of projectors are specified for each angular momentum.

The transformation is applied inside an augmentation-sphere of radius $r_{\rm PAW}$ centred on each atom. By construction, the all-electron and pseudo-partial waves are identical outside this region, such that $\ket{\phi_i^a} = \ket{\tilde{\phi_i^a}}$ for $|\mathbf{r} - \mathbf{R}_a| > r_{\rm PAW}$. Larger augmentation-spheres can reduce computational cost, but may compromise the accuracy of the reconstruction in regions where the wavefunction varies rapidly.

Expectation values of operators can be directly evaluated using the PAW transformation in Eqn.~\eqref{eqn:paw_transformation}, making it
sufficient to evolve pseudo-wavefunctions in time according to the PAW-transformed time-dependent Kohn-Sham equations

\begin{equation}
\label{eqn:paw_transformed_KS_equations}
i \hbar \tilde{S} \frac{\partial \ket{\tilde{\psi}_n}}{\partial t}
=
\tilde{\mathcal{H}}_{\rm PAW} \ket{\tilde{\psi}_n},
\end{equation}

\noindent
where $\tilde{S} = \hat{\mathcal{T}}^{\dagger} \hat{\mathcal{T}}$ is the PAW overlap operator, which accounts for non-orthogonality of the basis set. Increasing the number of projector functions improves the completeness of the basis set and so the accuracy of the all-electron reconstruction, but excessive completeness can introduce near linear-dependence within the basis set, leading to ill-conditioning of the generalised eigenvalue problem. Therefore, the number and energy range of projectors must be chosen to balance basis completeness with numerical stability.

\subsection{\label{subsec:dataset_generation_theory}Projector Augmented-Wave Dataset Generation}

PAW datasets are generated from reference all-electron atomic calculations. First, an atomic all-electron calculation is performed for a reference electronic configuration. Solving the radial Schr\"odinger equation, or alternatively a scalar-relativistic formulation \cite{koelling_technique_1977}, yields a set of all-electron radial wavefunctions. From these all-electron solutions, a set of all-electron partial-waves, $\ket{\phi_i}$, is selected (atom index is omitted since only a single atom is considered). Within the frozen-core approximation, electronic states designated as core-states are excluded from the set of all-electron partial-waves and are instead represented by a fixed charge-density. The choice and number of partial-waves is application-dependent, and represents a trade-off between accuracy and computational efficiency.

For each all-electron partial-wave, a corresponding pseudo partial-wave, $\ket{\tilde{\phi}_i}$, is constructed by fitting the all-electron radial function with a smooth analytic form (commonly a polynomial or spherical Bessel expansion). The pseudo and all-electron partial-waves are equal by construction beyond a chosen cut-off radius, $r_{\rm cut}$. Projector functions are then defined for each partial-wave. These projectors are constructed to be dual to the pseudo partial-waves, satisfying
\begin{equation} \label{eqn:biorthogonality_condition}
    \langle \tilde{p}_j| \tilde{\phi_k} \rangle = \delta_{jk},
\end{equation}
\noindent
and encode the projection of the pseudo wavefunctions onto their all-electron counterparts. In addition to partial-waves corresponding to bound states, scattering projectors may be added to extend completeness of the basis to higher energies. Differences between all-electron and pseudo partial-waves are used to define augmentation charge densities and operator corrections.

\section{\label{sec:methods}Methods}
\subsection{\label{sec:ehrenfest_dynamics_simulation}Stopping Power Calculations from Ehrenfest Dynamics Simulations}

Ehrenfest dynamics is implemented in the GPAW DFT code using a real-space wavefunction representation \cite{enkovaara_electronic_2010, ojanpera_nonadiabatic_2012}. Simulations were performed using the (adiabatic) local density approximation exchange-correlation functional provided by the LibXC library \cite{marques_libxc:_2012}. A ground-state DFT calculation is used as the initial condition for the TDDFT. The initial condition consists of a supercell expansion of FCC aluminium (lattice constant \SI{4.05}{\angstrom}) and a stationary proton placed at the starting position of the trajectory. When the projectile velocity is imposed at the first timestep, a short transient occurs in which the projectile charge state adapts to the new dynamical conditions. As noted in previous TDDFT stopping studies, this transient period can be excluded when extracting stopping powers \cite{kononov_reproducibility_2024, lee_multiscale_2020, schleife_accurate_2015}.

The proton projectile is described using GPAW's all-electron hydrogen setup, which replaces the PAW description with a regularised Coulomb potential, and contains no projectors or pseudised quantities. Consequently, there is no frozen-core approximation or all-electron reconstruction for the projectile, ensuring that any sensitivity to PAW datasets originates solely from the aluminium description.

Finite-size effects arise from the use of periodic boundary conditions, which allow the projectile to re-enter the simulation box and interact with electronic excitations generated earlier along the trajectory. It was found by Kononov et al.\ \cite{kononov_trajectory_2023} that core-electron contributions to stopping powers are converged when the minimum perpendicular distance between periodic images of a trajectory exceeds the Wigner-Seitz radius of the target material. Valence contributions were found to converge when this distance exceeded \SI{3}{\angstrom}. This empirical result is used to determine the size of the supercell expansion. For channelling trajectories, where such separation is more difficult to achieve, elongated supercells are used to maximise the propagation length before re-entry occurs.

Where off-channelling trajectories are simulated, the trajectory pre-sampling method developed by Gu et al.\ \cite{gu_efficient_2020} and Kononov et al.\ \cite{kononov_trajectory_2023} is used to select trajectories. This enables direct comparison of calculated stopping powers with experimental measurements and SRIM predictions whilst minimising computational cost.

In line with the findings of Schleife et al.\ \cite{schleife_accurate_2015} and Kononov et al.\ \cite{kononov_trajectory_2023}, stopping powers are extracted from simulations by a linear fit to projectile kinetic energy as a function of distance travelled along the trajectory.

Convergence tests were performed with respect to the real-space grid-spacing, timestep and numerical electron temperature. Simulations were performed using a grid spacing of approximately $h = 0.14\,$\SI{}{\angstrom}, with the number of grid-points along each direction chosen to optimise domain decomposition. Time-propagation was performed with a timestep corresponding to projectile displacement of $\Delta x = 0.04\,$\SI{}{\angstrom} per step. Electronic occupations were determined using a Fermi-Dirac distribution with an electronic temperature of \SI{0.1}{eV}, chosen to improve stability of time-propagation. All simulations were performed using only the $\Gamma$-point ($k=0$) for Brillouin-zone sampling.

\subsection{\label{subsec:paw_dataset_generation}PAW Dataset Generation}

Projector augmented-wave datasets for aluminium were generated using the functionality of GPAW. All aluminium datasets include $n=2$ electronic states in the valence partition rather than in the frozen core. This is motivated by previous TDDFT stopping power studies, where it was found that excitations of core electrons of the target material contribute significantly to stopping powers in cases where close approaches to host nuclei are simulated \cite{schleife_accurate_2015, correa_calculating_2018}. The $n=1$ states are retained in the frozen core, as generating pseudo partial-waves for these tightly bound states would require extremely small cut-off radii and high spatial resolution. All datasets were generated using a scalar-relativistic wave-equation. The GPAW PAW dataset generation tool allows the user to specify cut-off radii specific to the orbital angular momentum quantum number ($\ell$-channel). The maximum of these radii defines the augmentation radius $r_{\rm PAW}$.

\subsection{\label{subsec:dataset_generation_workflow}Dataset Generation Workflow}

Generating PAW datasets suitable for stopping power simulations requires a systematic procedure that balances basis completeness, numerical robustness, and transferability. In particular, the partial wave cut-off radii and projector construction must be chosen with care in order to meet these criteria. A greedy iterative workflow is adopted, in which augmentation and cut-off radii are first fixed, after which the basis is progressively refined through incremental extensions to the projector construction. Here, the greedy strategy refers to making a sequence of locally optimal choices, chosen because it allows the influence of each additional projector to be assessed in isolation, avoiding the combinatorial complexity that would arise from simultaneously optimising many parameters. Each intermediate dataset is validated against the criteria outlined in Section~\ref{subsec:dataset_tests}, allowing controlled and systematic convergence toward a dataset optimised for stopping power simulations.

In practice, the greedy character of the workflow is imperfect. The bi-orthogonality condition given by Eqn.~\eqref{eqn:biorthogonality_condition} means that introducing an additional projector modifies existing projector functions. Consequently, changes to one part of the basis inevitably lead to small adjustments throughout the projector set. In practice, these modifications are minor, and the workflow still provides a systematic route toward improved basis completeness.

The first step of the workflow is to identify the smallest augmentation radius and $\ell$-channel-specific cut-off radii compatible with stable ground-state calculations and acceptable logarithmic derivative agreement. The decision to use the smallest radii is justified in Section~\ref{subec:target_material_datasets}. With the augmentation radius and cut-off radii fixed, an initial minimal projector set is identified as a stable baseline from which the basis can be systematically extended.

The $\ell$-channel contributing most significantly to basis incompleteness is identified by generating candidate datasets in which a single additional projector is added to each $\ell$-channel. Stopping power simulations along a short trajectory are then performed for each candidate, and the resulting change in stopping power is used to quantify sensitivity to improvements in that $\ell$-channel. The $\ell$-channel producing the most significant change in stopping power is prioritised  for further refinement. This procedure ensures that basis-set extension is guided directly by non-adiabatic processes, rather than by ground-state considerations alone. These simulations were performed for protons with initial kinetic energy of \SI{400}{keV}, since high projectile energies probe both core and valence electronic excitations and therefore provide a stringent test of basis completeness.

Once the dominant channel has been identified, a single projector is added to that channel. The projector energy is then adjusted to maximise computational efficiency while preserving the improvement in stopping power predictions. Computational efficiency provides a practical proxy for the conditioning of the time-dependent equations, since overlap within the projector construction leads directly to slower convergence during time-propagation. This optimisation step improves numerical robustness and can enable the inclusion of additional projectors before conditioning-related solver stagnation is encountered. 

\subsection{\label{subsec:dataset_tests}Dataset Requirements}

Each dataset was evaluated using a three-part validation procedure. The first step is an inspection of logarithmic derivatives of the all-electron and pseudo partial-waves and the corresponding energy-integrated deviation. Good agreement ensures that the dataset reproduces the correct phase-shifts for electron scattering. This test also reveals ghost states, which arise from unphysical solutions of the pseudo-Hamiltonian and manifest as discontinuities in the logarithmic derivative curves.

The second validation step is an equation of state (EOS) test. This evaluates the ground-state energy as a function of volume, from which the equilibrium lattice constant and bulk modulus are obtained from a third-order Birch-Murnaghan fit. Agreement with reference all-electron calculations indicates that the dataset accurately reproduces the bonding characteristics and elastic response of the material. EOS comparisons constitute standard practice in the validation of pseudopotentials and PAW datasets \cite{bosoni_how_2023, jolletGenerationProjectorAugmentedWave2014}.

Finally, short time-dependent stopping simulations provide a practical assessment of dataset performance under non-equilibrium conditions. Numerical stability of the propagator is monitored, which can reveal issues such as over-completeness of the projector construction. These simulations were performed on a short \SI{36}{\angstrom} pre-sampled trajectory. While insufficient to reproduce the full statistical distribution of impact parameters sampled in experimental measurements, the chosen trajectory probes a range of impact parameters to nuclei.

\section{\label{sec:results}Results}
\subsection{\label{subec:target_material_datasets}Target Material Datasets}

\subsubsection{\label{subsubsec:radii_results}Influence of PAW Dataset Radii}

\begin{figure}[!t]
\includegraphics[width=\columnwidth]{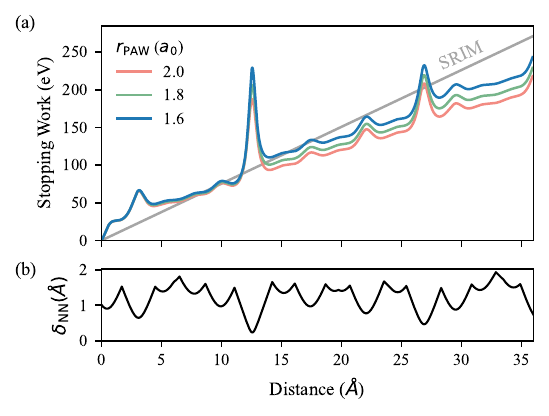}
\caption{\label{fig:dataset_radius}a) Cumulative stopping work for a proton projectile with initial kinetic energy of \SI{400}{keV} on a \SI{36}{\angstrom} off-channelling trajectory using aluminium PAW datasets with different augmentation radii $r_{\rm PAW}$. Starting from the baseline dataset listed in Table~\ref{tab:projector_datasets}, cut-off radii for all $\ell$-channels are uniformly increased in \SI{0.2}{\bohr} increments. The augmentation radius $r_{\rm PAW}$ increases accordingly. The grey line shows the stopping power predicted by SRIM, plotted as the cumulative work assuming a constant stopping force. b) shows the nearest neighbour distance from the projectile to aluminium nuclei along the trajectory.}
\end{figure}

The influence of the augmentation radius and cut-off radii is first examined. Fig.~\ref{fig:dataset_radius} shows the cumulative work done on a proton projectile with an initial kinetic energy of \SI{400}{keV} travelling along a short off-channelling trajectory using PAW datasets with varying augmentation and cut-off radii. Starting from dataset 0 listed in Table~\ref{tab:projector_datasets}, the cut-off radii for all angular momentum channels are uniformly increased in increments of \SI{0.2}{\bohr}. Since the augmentation radius $r_{\rm PAW}$ is defined as the largest $\ell$-channel cut-off radius, it therefore increases accordingly.

The plot in Fig.\ref{fig:dataset_radius} shows cumulative stopping work done on the projectile, which is defined as the instantaneous kinetic energy relative to its initial value. Stopping power, $S = -dE/dx$, corresponds to the average slope of the cumulative stopping work as a function of distance travelled. The peaks in stopping work correspond to close-passes of the projectile with aluminium nuclei, where the projectile interacts with the steep Coulomb potentials of nuclei and directly probes core electronic density. Between the peaks, the more gradual increase in stopping work is due to the interaction with valence electrons.

It can be seen from the plot in Fig.~\ref{fig:dataset_radius} that smaller augmentation and cut-off radii lead to increased stopping power, showing improved accuracy when compared to the SRIM prediction. Smaller cut-off radii confine the pseudo-wavefunctions to a smaller region around each nucleus. This improves the reconstruction of the rapidly varying all-electron wavefunctions near the nucleus, leading to improved predictions of electronic stopping. 

These datasets use a minimal projector construction, and as such, significantly underestimate stopping power for high-energy projectiles traversing off-channelling trajectories. Nevertheless, the observed improvement in stopping power with decreasing PAW radius motivates the use of the minimum stable radii in the dataset generation workflow adopted throughout the remainder of this work.

\subsubsection{\label{subsubsec:projector_results}Influence of Projector Construction}

Next, the sensitivity of stopping power predictions to the projector construction is investigated. Datasets were generated using the workflow described in Section~\ref{subsec:dataset_generation_workflow}. Key parameters of the identified datasets are shown in Table~\ref{tab:projector_datasets}. Within each $\ell$-channel, projectors were added at progressively higher reference energies, a pattern that emerged naturally during the optimisation process. Significant improvements to stopping predictions were most often obtained by specifying additional projectors with angular momentum $\ell = 2$. This highlights the importance of accurately describing high-energy scattering states with $\ell = 2$ near the projectile. Datasets were generated using the workflow until the defined requirements could no longer be met. The first failure mode encountered was projector over-completeness, which leads to poor conditioning and convergence failures during time-propagation.

\begin{figure}[!t]
\includegraphics[width=\columnwidth]{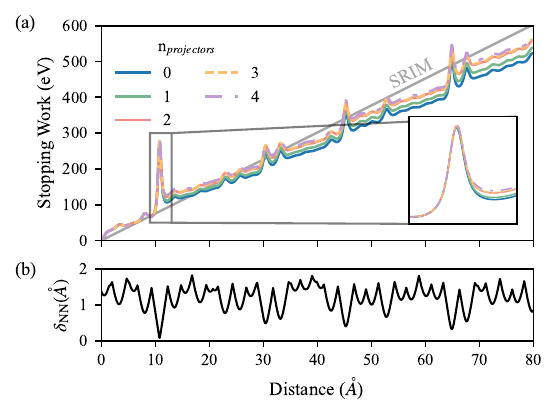}
\caption{\label{fig:dataset_projectors}a) Cumulative stopping work for a proton projectile with initial kinetic energy of \SI{400}{keV} on a \SI{80}{\angstrom} off-channelling trajectory using aluminium PAW datasets that systematically extend basis completeness through the proposed iterative workflow. The grey line shows the stopping power predicted by SRIM, plotted as the cumulative work assuming a constant stopping force. The inset axes show the stopping work in the region of a close-approach with an aluminium nucleus. b) shows the nearest neighbour distance from the projectile to aluminium nuclei along the trajectory.}
\end{figure}

Fig.~\ref{fig:dataset_projectors} shows cumulative stopping work done on a proton with initial kinetic energy of \SI{400}{keV}. Each trace corresponds to an aluminium dataset in Table~\ref{tab:projector_datasets}. The first two additional projectors produce substantial improvements in the predicted stopping. Subsequent projectors provide diminishing improvements while increasing computational cost.

The largest differences between stopping predictions with different datasets occur around the sharp peaks in stopping work, when the projectile passes close to host nuclei in regions approximated with PAW. This behaviour is highlighted in the inset of Fig.~\ref{fig:dataset_projectors}, which shows a representative close approach where increasing the completeness of the projector basis systematically increases the predicted stopping work. Consequently, the minimum impact parameter sampled by the projectile is an important factor in determining the required projector construction. Between the peaks, the projectile is further from host nuclei, meaning that interactions occur largely outside the augmentation spheres, where pseudo and all-electron wavefunctions coincide by construction. Here, stopping is therefore largely insensitive to the projector construction.

\begin{table*}
\caption{\label{tab:projector_datasets}PAW datasets generated using the workflow described. All datasets contain projectors for the bound-states included in the valence partition; only additional scattering projectors are listed.}
\begin{ruledtabular}
\begin{tabular}{cccccccc}
 & & \multicolumn{3}{c}{Cut-off radii (\SI{}{\bohr})} & \multicolumn{3}{c}{Projector energies (Ha)}\\
 Dataset number & $r_{\rm PAW}$ (\SI{}{\bohr}) & s & p & d & s & p & d\\
\hline
 0 & 1.6 & 1.2 & 1.5 & 1.6 & -- & -- & -- \\
 1 & 1.6 & 1.2 & 1.5 & 1.6 & -- & -- & 0.5 \\
 2 & 1.6 & 1.2 & 1.5 & 1.6 & -- & -- & 0.5, 4.0 \\
 3 & 1.6 & 1.2 & 1.5 & 1.6 & -- & 3.0 & 0.5, 4.0 \\
 4 & 1.6 & 1.2 & 1.5 & 1.6 & -- & 3.0 & 0.5, 4.0, 6.5 \\
\end{tabular}
\end{ruledtabular}
\end{table*}

\subsubsection{\label{subsec:projector_protocol}Dataset Selection Protocol}

Since differences between PAW datasets originate from their ability to represent electronic excitations during close projectile-nucleus encounters, the optimal dataset depends on the impact parameters and projectile energies sampled by a particular simulation. Consequently, dataset selection should be regarded as an application-specific convergence problem rather than a fixed methodological choice.

To illustrate such a convergence procedure, a proton with an initial kinetic energy of \SI{400}{keV} was propagated along a trajectory designed to produce six successive close approaches with progressively decreasing impact parameters. For each close encounter, $i$, the collision-resolved stopping work, $W_i^n$, is defined as the increase in cumulative stopping work over a fixed window centred on the minimum projectile-nucleus separation. This is done for each dataset, labelled $n$.

To quantify the extent to which a given PAW dataset reproduces the stopping associated with an individual collision, the collision-resolved stopping work is normalised by that obtained by the most complete numerically stable dataset considered in this work. This defines the relative stopping fraction,
\begin{equation}
    \zeta_i^{n} = \frac{W_i^{n}}{W_i^{N}}
\end{equation}
\noindent
where $N$ is the most complete numerically stable dataset considered (dataset 4 in Table~\ref{tab:projector_datasets}). For the purposes of this work, a dataset is considered sufficient for the desired accuracy for a given impact parameter when $\zeta_i^n \ge 0.95$. A threshold of 95\% is adopted here as a representative compromise between computational efficiency and accuracy, although the same protocol may be applied using any application-appropriate stopping-fraction criterion.

\begin{figure}[!t]
\includegraphics[width=\columnwidth]{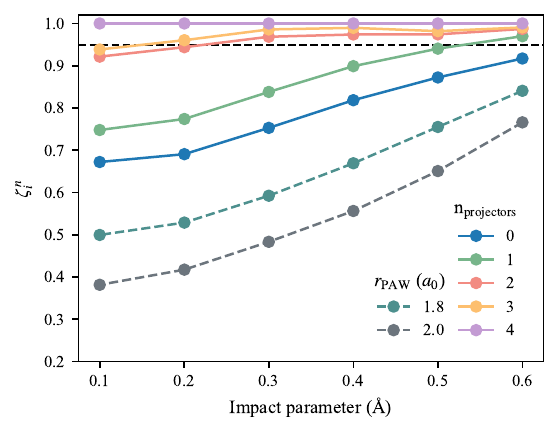}
\caption{\label{fig:projector_protocol_temp}Relative stopping fraction, $\zeta_i^n$, as a function of impact parameter for the five PAW datasets listed in Table~\ref{tab:projector_datasets}. Additional traces corresponding to datasets with larger PAW radii are included for completeness. The stopping fraction is defined relative to the most complete numerically stable dataset (dataset 4). The horizontal dashed line indicates the 95\% criterion adopted for dataset selection; the intersection of each curve with this threshold defines the smallest impact parameter for which that dataset is considered adequate.}
\end{figure}

Fig.~\ref{fig:projector_protocol_temp} shows the relative stopping fraction as a function of impact parameter for the five PAW datasets considered. As expected, datasets recover a greater fraction of the stopping predicted by the reference calculation at large impact parameters, while progressively more complete projector constructions are required as the projectile penetrates further into the PAW augmentation spheres. The intersection of each curve with the 95\% threshold therefore defines the smallest impact parameter for which that dataset satisfies the chosen accuracy criterion. In practice, only a small number of short calibration simulations are required to determine the least computationally expensive dataset appropriate for the projectile energies and impact parameters expected in subsequent production calculations.

Applying this protocol to the projectile energies and impact parameters considered in the present work indicates that the dataset containing three additional scattering projectors provides a near-converged description of stopping while remaining substantially less computationally demanding than dataset 4. Although dataset 3 falls marginally below the 95\% stopping-fraction criterion, these events are infrequent and so the resulting differences in the total stopping are expected to be small. Dataset 3 is therefore adopted for all subsequent stopping simulations unless otherwise stated.

An additional advantage of this approach is that many PAW implementations, including GPAW, permit different atoms of the same chemical species to employ different PAW datasets. Consequently, only atoms expected to experience close projectile encounters need to be described using the most complete datasets, while more distant atoms may be represented using computationally cheaper datasets.

\subsection{\label{subsec:off_channelling_stopping_powers}Off-Channelling Stopping Powers}

\subsubsection{\label{subsubsec:velocity_dependence_results}Velocity Dependence}

The velocity dependence of electronic stopping power is examined using a pre-sampled off-channelling trajectory. Meaningful comparison with SRIM requires careful trajectory selection, since core-electron response contributes significantly to the total stopping power. If the sampled impact parameters do not resemble those encountered in an experimental measurement with a polycrystalline sample, comparisons with models assuming homogeneous materials become unreliable. The trajectory used here satisfies the sampling criterion proposed by Kononov \cite{kononov_trajectory_2023}, achieving a trajectory metric of $D_H = 0.06$ after \SI{80}{\angstrom}, where $D_H$ is the Hellinger distance to the nearest-neighbour sampling distribution of an infinitely-long trajectory. Kononov suggested that representative sampling of the supercell is achieved for $D_H < 0.1$.

\begin{figure}[!t]
\includegraphics[width=\columnwidth]{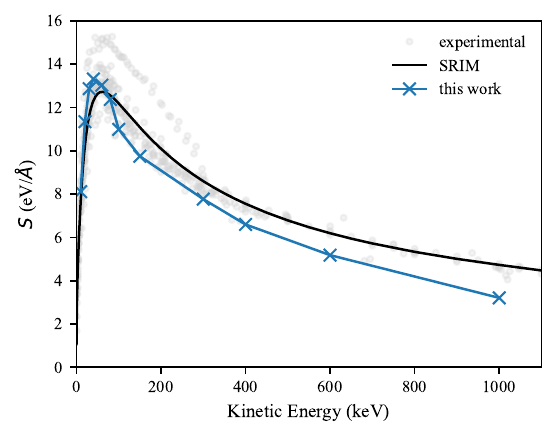}
\caption{\label{fig:stopping_curve}Proton stopping power in aluminium as a function of projectile kinetic energy obtained from TDDFT Ehrenfest dynamics simulations. SRIM predictions and experimental data \cite{noauthor_electronic_2023} is shown for comparison.}
\end{figure}

Fig.~\ref{fig:stopping_curve} shows the velocity dependence of proton stopping power, with a comparison to experimental data \cite{noauthor_electronic_2023} and to SRIM. Agreement with SRIM is excellent below the Bragg peak. At higher projectile energies, the calculated stopping powers are systematically lower than SRIM predictions. This behaviour indicates that the dataset optimisation procedure captures the dominant physics governing stopping below the Bragg peak, while discrepancies at higher energies likely originate from methodological limitations, which are addressed below, rather than deficiencies in the PAW construction.

\subsubsection{\label{subsubsec:remaining_sources_of_error}Remaining Sources of Error}

Several sources of error remain that cannot be eliminated through PAW dataset refinement. These arise from approximations inherent to the simulation methodology and from physical processes that are not fully captured within the present approach.

First, the frozen-core approximation prevents excitation of electrons belonging to the frozen-core. In the datasets considered, the frozen core included $n=1$ states, whose contribution to stopping increases with projectile energy. Treating these states explicitly would require PAW datasets in which $n=1$ electrons are included in the valence partition. In practice this requires extremely small partial-wave cut-off radii, making such datasets difficult to construct within the GPAW framework. Real-time TDDFT simulations of proton stopping in liquid water using pseudopotentials that explicitly include the $n=1$ electrons in the valence partition have shown that $n=1$ excitations contribute negligibly below projectile kinetic energies of approximately \SI{50}{keV}, but can increase the stopping power by $20$–$30\%$ at much higher energies \cite{yao_k_2019}. These studies also demonstrate that core excitations modify the valence response through shake-up processes, meaning that stopping power cannot be decomposed into independent valence and core contributions. Although the contributions in aluminium may differ, these results suggest that neglect of deep-core excitations is likely to be the primary contributor to the remaining underestimation of stopping power at projectile energies above the Bragg peak.

A second limitation arises from the finite supercell expansion. Collective electronic excitations such as plasmons are captured in the Hartree potential term of the Kohn-Sham Hamiltonian, but the  wavelengths of excitations are limited by the size of the simulation when combined with periodic boundary conditions. As discussed by Correa \cite{correa_calculating_2018}, this leads to a systematic underestimation of stopping power at sufficiently high projectile energies. Approximate corrections based on calculations using the Lindhard dielectric model have been proposed \cite{schleife_accurate_2015}.

Additional uncertainty arises from the use of Ehrenfest dynamics. In these simulations, the projectile velocity evolves continuously as energy is transferred to the electronic subsystem, meaning that the projectile may lose a non-negligible fraction of its kinetic energy along the simulated trajectory. As a result, the stopping power varies with time as the projectile slows. This differs from conventional constant-velocity TDDFT simulations, in which the stopping power is evaluated at a fixed projectile kinetic energy. The effect is more pronounced at low energies, where stopping power varies strongly with velocity and the relative energy loss along the trajectory is larger. Consequently, the stopping power extracted from an Ehrenfest dynamics simulation represents an average over a range of projectile energies.

\subsection{\label{subsec:channelling_stopping_powers}Channelling Stopping Powers}

\begin{figure*}[htbp]
\includegraphics[width=\textwidth]{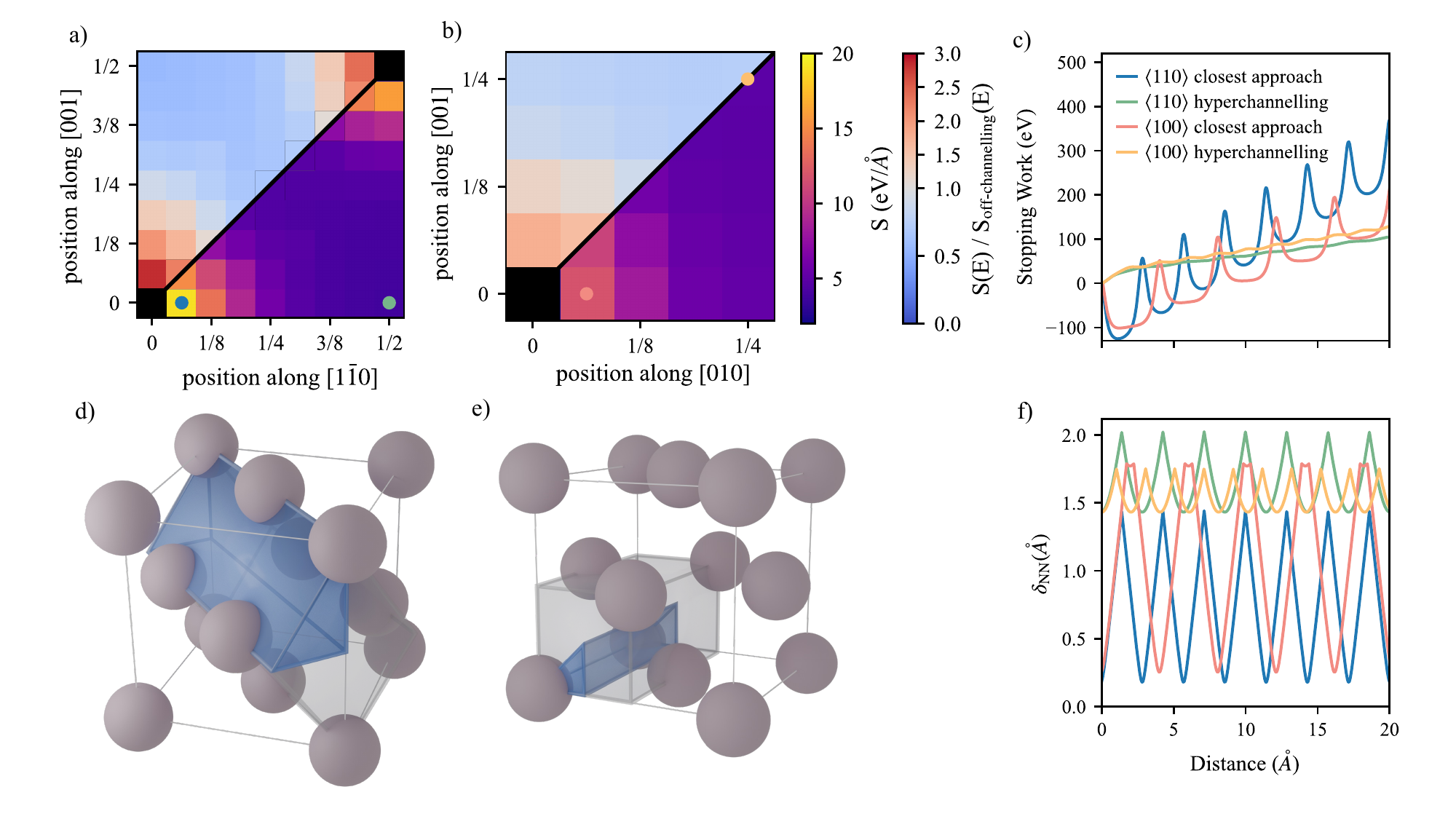}
\caption{\label{fig:channelling}Stopping power of protons with initial kinetic energy of \SI{400}{keV} travelling along a) $
\langle 110 \rangle$ and b) $
\langle 100 \rangle$ directions. The projectile starting position is specified inside the symmetry-irreducible section of the plane perpendicular to the direction of travel. The coordinates axes are normalised by the lattice repeat distance along the corresponding crystallographic directions ($a$ for $\left[ 001 \right]$ and $\left[ 010 \right]$, and $a/\sqrt{(2)}$ for $\left[ 1\bar{1}0 \right]$). Trajectories through lattice points are excluded from sampling and are shown in black. Values below the diagonal show the raw stopping power in \SI{}{eV / \angstrom}, while values above the diagonal show the stopping power normalised to the simulated off-channelling value at the same energy. c) shows cumulative stopping work for the selected trajectories in panels a and b, shown by the coloured circles. Diagrams of the FCC unit cell, with the d) $
\langle 110 \rangle$ and e) $
\langle 100 \rangle$ channel shown in grey, and the symmetry-irreducible section of the channel shown in blue. f) shows nearest neighbour distances to nuclei along the selected trajectories.}
\end{figure*}

Having established good accuracy for off-channelling trajectories, channelling effects in FCC aluminium are now investigated. In channelling configurations the projectile propagates along low-index lattice directions, providing a regime in which the periodic lattice potential plays a central role in determining projectile dynamics. These effects underpin a range of applications involving the control of charged-particle beams, including beam steering and collimation using bent crystals \cite{redaelli_crystal_2025}, as well as proposed `super-focussing' effects \cite{demkov_channeling_2009}. These phenomena are often modelled using Monte Carlo approaches \cite{cai_simulation_2024}. Ehrenfest dynamics simulations provide a route to directly capture these effects from first-principles.

Fig.~\ref{fig:channelling} shows stopping powers for \SI{400}{keV} projectiles travelling along $\langle 110 \rangle$ and $\langle 100 \rangle$ channelling trajectories for a range of initial positions in the plane perpendicular to the direction of propagation. Stopping powers are presented in two forms. Raw stopping power, in units of \SI{}{eV/\angstrom}, is shown below the diagonal. Stopping power as a fraction of the off-channelling stopping power prediction is shown above the diagonal. Trajectories through lattice points are not simulated, and are shown as black pixels.

For both channelling directions, stopping power varies significantly across the channel cross-section. Trajectories passing close to lattice sites correspond to small impact parameters with respect to nuclei and repeatedly probe regions of high electronic density. These trajectories exhibit stopping powers that exceed the off-channelling value. Conversely, trajectories confined to the central regions of the channel interact predominantly with valence electrons and exhibit reduced stopping. In both $\langle 100 \rangle$ and $\langle 110 \rangle$ channels, the minimum stopping power occurs for trajectories passing through the geometric centre of the channel (hyperchannelling trajectories). As shown in Fig.~\ref{fig:channelling}, these trajectories have identical impact parameters, but traverse distinct crystallographic sites (tetrahedral for $\langle 100 \rangle$ and octahedral for $\langle 110 \rangle$).

Despite the identical impact parameters, these two trajectories exhibit different stopping powers. This difference arises from the periodicity of the crystal along the direction of propagation. For the $\langle 100 \rangle$ channel, symmetry reduces the effective periodicity of the hyperchannelling trajectory to $a/2$, whereas for $\langle 110 \rangle$, it remains $a/\sqrt2$, where $a$ is the lattice parameter. As a result, even when the impact parameter is the same, the projectile in the $\langle 100 \rangle$ channel samples higher-density regions more frequently. The stopping power therefore reflects not only the maximum density encountered, but also the spatial frequency of such encounters along the trajectory. A further consequence of the channel geometry is reflected in the overall spread of stopping powers. The $\langle 100 \rangle$ channel exhibits smaller variation in stopping powers compared to the $\langle 110 \rangle$ channel, consistent with the higher symmetry of its hyperchannelling trajectory.

This interplay between impact parameter and sampling frequency also governs the behaviour of trajectories with the highest stopping powers. Care must first be taken when comparing trajectories along different channels in Fig.~\ref{fig:channelling}, as identical fractional coordinates correspond to different physical impact parameters due to the distinct transverse length scales. However, this ambiguity can be removed by directly comparing trajectories with equal impact parameters. For example, the trajectory at $(1/2,7/16)$ in the $\langle 110 \rangle$ channel has an equal impact parameter to the $(1/16, 0)$ trajectory in the $\langle 100 \rangle$ channel. The higher stopping power of the $\langle 110 \rangle$ trajectory can be attributed primarily to the higher frequency with which close encounters occur. This effect is enhanced by the fact that the $\langle 100 \rangle$ reaches larger distances from lattice sites between successive close approaches. These results reinforce that stopping is governed not only by the magnitude of the closest approach, but also by the periodicity with which such interactions are sampled along the trajectory.

The present results are obtained at sufficiently high projectile energies that deflection by the lattice potential is negligible over the simulated distances, allowing the role of electronic stopping along well-defined trajectories to be isolated. At lower energies, the coupling between projectile motion and the lattice potential gives rise to transverse oscillations within the channel. The stopping power associated with such motion would therefore represent an average over a range of impact parameters. Capturing behaviour in this regime represents a natural extension of the present work due to the capability of Ehrenfest dynamics to self-consistently treat ion dynamics.

\section{\label{sec:conclusion}Conclusion}

A systematic approach to the generation of projector augmented-wave (PAW) datasets for time-dependent density functional theory simulations of charged-particle stopping powers has been developed. Results show that PAW datasets optimised for ground-state calculations are not necessarily sufficient for stopping simulations. Accurate predictions require careful control of augmentation radii, cut-off radii, and the projector construction to capture non-adiabatic effects while maintaining numerical robustness. Sensitivity to dataset design depends strongly on the physical regime, with high-energy projectiles and small impact parameters requiring an accurate description of core-electron response, whereas valence-dominated regimes are less demanding. Building on these observations, a collision-resolved convergence protocol has been introduced that relates dataset completeness directly to the impact parameters and projectile velocities sampled during a simulation. This provides a transferable procedure for selecting the least computationally expensive PAW dataset appropriate for a given application.

The methodology introduced here provides a general framework for constructing PAW datasets suitable for non-adiabatic simulations and is readily applicable to other materials. Remaining limitations of the PAW datasets arise primarily from the frozen-core approximation, which neglects excitations of deeply bound states that become important at high projectile energies. Addressing this would require the development of PAW datasets which explicitly treat all core states, which remains a direction for future work. In addition, extending simulations to lower-energy regimes, where dynamically evolving channelling trajectories can be treated within first-principles simulations represents a promising avenue for further research.

\begin{acknowledgments}
The work received support from EPSRC and First Light Fusion under the AMPLIFI Prosperity partnership, grant no. EP/X025 373/1. The author is grateful for the use of computing resources provided by STFC Scientific Computing Department’s SCARF cluster. 
\end{acknowledgments}

\bibliography{citations}

\end{document}